\documentclass[runningheads]{llncs}
\usepackage[T1]{fontenc}
\usepackage{graphicx}

\usepackage{amssymb}
\usepackage{amsmath}
\usepackage{xcolor}
\usepackage{siunitx}
\usepackage{tikz}
\usepackage{pgfplots}
\usepgfplotslibrary{groupplots}
\usetikzlibrary{spy,decorations.fractals,shapes.misc}
\usetikzlibrary{decorations.markings}
\usepgfplotslibrary{colormaps}
\usepgfplotslibrary{colorbrewer}
\pgfplotsset{cycle list/Dark2}
\pgfplotsset{compat=1.18}
\usepackage{subcaption}

\begin{document}
\title{Modeling of Chemically Reactive Gas Flows for Multi-Scale Problems}
%
%
\author{Franziska Tuttas\orcidID{0000-0003-3898-0908} \and
Marcel Pfeiffer\orcidID{0000-0002-4087-1789}}
\authorrunning{F. Tuttas et al.}
%
\institute{Institute of Space Systems, University of Stuttgart, Pfaffenwaldring 29, 70569 Stuttgart, Germany\\
\email{tuttasf@irs.uni-stuttgart.de}}
\maketitle              
\begin{abstract}

Accurate modeling of chemical reactions plays a key role in the investigation of high-temperature flows.
The Direct Simulation Monte Carlo (DSMC) method provides highly accurate solutions for such flows.
However, its computational cost becomes prohibitive for low Knudsen regimes.
Particle-based continuum methods employing the Bhatnagar-Gross-Krook (BGK) or Fokker-Planck approximations have gained increasing attention in recent years, improve computational efficiency in low-Knudsen regimes by modeling relaxation processes rather than binary collisions, while retaining the particle approach that enables straightforward coupling with DSMC.
In the last years, different continuum-based models in the open-source particle code PICLas have been extended to simulate gas mixtures with internal energy modes.
The present work focuses on modeling of chemically reactive gas mixtures in the Ellipsoidal Statistical BGK model.
Model accuracy is assessed for two test cases of a reactive gases confined in a reservoir.

\end{abstract}

\section{Introduction}
\label{sec:intro}
Numerical results for flows over a wide range of Knudsen numbers are of utmost importance to gain a deeper understanding of the underlying physical processes, in particular due to limitations in experimental investigations.
The well-established Direct Simulation Monte Carlo (DSMC) method~\cite{bird} is a highly reliable approach to simulate non-equilibrium gas flows, proven to produce accurate results.
However, since the computational costs of DSMC increase significantly in the low Knudsen regimes, alternative approaches are required.
In recent years, particle-based continuum methods, e. g. the Bhatnagar-Gross-Krook (BGK)~\cite{bgk} or Fokker-Planck (FP)~\cite{fp-basic} approximations, have gained increasing attention for simulating multi-scale problems, as they improve computational efficiency in low Knudsen regimes.
Due to the same particle approach, a simple coupling with the DSMC method is possible, where DSMC is then used for the high Knudsen flow regions and BGK or FP for the lower ones.
The BGK and FP methods are based on a relaxation process of the particle distribution function towards a specific target distribution instead of modeling binary collisions, leading to less restrictions on the time step and simulation particle number compared to DSMC.

The particle-based Ellipsoidal Statistical BGK (ESBGK)~\cite{esbgk} model implemented in the open-source code PICLas~\cite{piclas,piclas-bgk} has previously been extended to simulate polyatomic gas mixtures~\cite{bgk-polymulti}.
It is advantageous compared to the standard BGK model, as it produces the correct Prandtl number of the simulated gas mixture due to the use of an anisotropic Gaussian distribution function.

This work focuses on the extension of the ESBGK model to include chemical reactions in gas mixtures, which is of great importance in particular for space applications such as spacecraft re-entry.
Here, the flow is characterized by high temperatures, where chemical reactions such as the dissociation of molecular oxygen and nitrogen cannot be neglected.
Previous work on the implementation of chemical reactions in the ESBGK model~\cite{bgk-chem} is updated and extended by a presentation of the chemistry model.
The theoretical basics of chemical reactions are briefly explained in section~\ref{sec:chemistry}, while the extension of the ESBGK model and the subsequent implementation in PICLas are discussed in sections~\ref{sec:esbgk} and \ref{sec:implementation}.
For verification purposes, two different test cases of a reactive gas mixture confined in a reservoir are presented to demonstrate the functionality of the proposed model.
The solutions of the simulations are compared to DSMC results in section~\ref{sec:results}.

\section{Theory}

\subsection{Chemical reactions}
\label{sec:chemistry}
An elementary reaction is defined as
\begin{equation}
    \sum_{s=1}^{M_\text{reac}} \nu_s' \text{A}_s \xrightarrow{k_\text{f}} \sum_{s=1}^{M_\text{prod}} \nu_s'' \text{A}_s + \Delta H_r
\end{equation}
with stoichiometric coefficients $\nu_s'$ of the reactants on the left-hand side and $\nu_s''$ of the products on the right-hand side, the forward rate constant $k_\text{f}$, and the heat of reaction $\Delta H_r$.
The latter is calculated for each reaction $r$ using the heat of formation $\Delta H_{\text{f}}$ of all reactants and products involved:
\begin{equation}
    \Delta H_r = \sum_{s=1}^{M_\text{prod}} \nu_s'' \Delta H_{\text{f},s} - \sum_{s=1}^{M_\text{reac}} \nu_s' \Delta H_{\text{f},s}.
\end{equation}
Species-specific vkalues for $\Delta H_{\text{f}}$ can be found within the Active Thermochemical Tables (ATcT)~\cite{ATcT}.
If energy is released during the reaction ($\Delta H_r < 0$), it is an exothermic reaction, if energy is required instead ($\Delta H_r < 0$), it is an endothermic reaction.
The rate constant is defined as
\begin{equation}
    k(T) = A T^b \exp\left(-\frac{E_\text{a}}{k_\text{b}T}\right)
\end{equation}
according to the extended Arrhenius equation, using reaction-specific constants $A$ and $b$ and activation energy $E_\text{a}$ to fit experimentally determined reaction rates.
With this, the change of the particle number densities of reactants and products over time can be described for a specific reaction:
\begin{equation}
    \frac{\partial n_s}{\partial t} = k \left(\nu_s''-\nu_s'\right) \prod_{k=1}^M n_k^{\nu_k'}.\label{eq:dens-change}
\end{equation}

\subsection{ESBGK chemistry model}
\label{sec:esbgk}

The BGK operator approximates the Boltzmann collision integral by a relaxation process of the particle distribution function $f_s \left(\mathbf x,\mathbf v, t \right)$ of each species $s$ with position $\mathbf{x}$ and velocity $\mathbf{v}$ at time $t$ towards a target distribution $f_s^t$ with a relaxation frequency $\nu$~\cite{bgk}:
\begin{align}
    \partial_t f_s + \mathbf {v} \partial_{\mathbf x} f_s &= \nu(f_s^t-f_s). \label{eq:bgk} \\
    f_{s,\text{Atom}}^t &= f_s^{t, \text{tr}}(\mathbf v), \\
	f_{s,\text{Molecule}}^t &= f_s^{t, \text{tr}}(\mathbf v)f_s^{t,\text{rot}}(E_\text{rot})f_s^{t,\text{vib}}(i_\text{vib}).
\end{align}
The target distribution function $f_s$ for an atomic species only consists of a translational part, while for a molecular species, it can be separated into the translational, the rotational and the vibrational part~\cite{MATHIAUD202265,dauvois2021bgk}
with the rotational and vibrational components of the distribution function taken directly from~\cite{MATHIAUD202265}.
Here, $E_\text{rot}$ is the rotational energy and $i_\text{vib}$ the vibrational quantum number.

The proposed ESBGK model is based on a combination of different models~\cite{MATHIAUD202265,energy-cons,brull,brull2021ellipsoidal}.
The correct Prandtl number is achieved with an anisotropic Gaussian distribution for the translational part of the target distribution function~\cite{esbgk}:
\begin{align}
	f_s^{\text{ESBGK,tr}} &= \frac{n_s}{\sqrt{\det 2\pi\mathbf A_s}} \exp\left[-\frac{1}{2}\mathbf c^\text{T} \mathbf A_s^{-1} \mathbf c\right], \label{eq:esbgk} \\
    \mathbf A_s &= \frac{k_\text{B} T_{\text{tr,rel}}}{m_s} \mathbf I - \frac{1-\alpha Pr}{\alpha Pr}\left(\frac{T_{\text{tr,rel}}}{T_{\text{tr}}} \mathbf{\Theta}- \frac{k_\text{B} T_{\text{tr,rel}}}{m_s}\mathbf I\right), \\
    \mathbf{\Theta} &= \frac{\mathbf{P}}{\rho} = \frac{1}{\rho} \sum_{s=1}^M m_s\int \mathbf c \mathbf c^\text{T} f_s\,\text d\mathbf v, \\
    \alpha &= m \frac{\sum_{s=1}^M \frac{n_s}{m_s} \left(5 + \xi_{\text{vib},s}+\xi_{\text{rot},s}\right)}{\sum_{s=1}^M n_s \left(5 + \xi_{\text{vib},s}+\xi_{\text{rot},s}\right)}, \quad m = \sum_{s=1}^M \frac{n_s}{n} m_s.
\end{align}
Here, $n_s$ is the number density of species $s$, $\mathbf c = \mathbf v - \mathbf u$ is the relative velocity of the particles with respect to the bulk velocity $\mathbf u$, and $\mathbf A_s$ is the anisotropic matrix, using the Boltzmann constant $k_\text{B}$, the molecular mass $m_s$, the targeted Prandtl number $Pr$, the translational temperature $T_{\text{tr}}$, and the relaxation translational temperature $T_{\text{tr,rel}}$, defined in~\cite{MATHIAUD202265}.
The symmetric stress tensor $\mathbf{\Theta}$ is calculated from the pressure tensor $\mathbf P$ with mass density $\rho$.
$\alpha$ is a model variable depending on particle number densities, mass $m$ and molecular masses, and internal degrees of freedom $\xi_s$~\cite{brull2021ellipsoidal}.
The relaxation frequency of the model is defined as
\begin{equation}
	\nu = \frac{nk_{\text{B}} T_{\text{tr}}}{\mu}\alpha Pr
\end{equation}
with the viscosity of the gas mixture $\mu$, which is calculated by using the first approximation to the viscosity depending on collision integrals~\cite{collint}.
The rotational and vibrational components of the distribution function are are taken directly from Ref.~\cite{MATHIAUD202265}:
\begin{alignat}{1}
	f_s^{\mathrm{ESBGK,rot}} = &\frac{E_\mathrm{rot}^{(\xi_{\mathrm{rot},s}-2)/2}}{\Gamma ( \xi_{\mathrm{rot},s}/2)\left(k_\mathrm{B} T_{\mathrm{rot, rel,s}}\right)^{\xi_{\mathrm{rot},s}/2}} \cdot\exp\left[-\frac{E_\mathrm{rot}}{k_\mathrm{B} T_{\mathrm{rot, rel,s}}}\right],\\
	f_s^{\mathrm{ESBGK,vib}} = &\prod_{j=1}^\gamma \left(1-\exp\left[-\frac{\Theta_{\mathrm{vib},j,s}}{T_{\mathrm{vib, rel},s}}\right]\right) \cdot\exp\left[-i_j\frac{\Theta_{\mathrm{vib},j,s}}{T_{\mathrm{vib, rel},s}}\right].
\end{alignat}
The relaxation temperatures $T_{\mathrm{tr, rel}}$, $T_{\mathrm{rot,rel},s}$ and $T_{\mathrm{vib,rel},s}$ introduced here are chosen so that the relaxation corresponds to the Landau-Teller relaxation.
The heat of reaction per time $\Delta H$ is distributed to the relaxation energies according to the degrees of freedom of the gas
\begin{alignat}{2}
    \left<E\right>_{r,s}(T_{r,\text{rel},s}) &= &&\,\left<E\right>_{r,s}(T_{r,s}) + \frac{\tau}{Z_{r,s}\tau_{s,\text{C}}}\left(\left<E\right>_{r,s}(T_{\text{tr}}) - \left<E\right>_{r,s}(T_{r,s}) \right) \nonumber \\
    & &&\,- \tau \Delta H \frac{\xi_{r,s} n_s}{3 n + \sum_r\sum_{s}^{M_\text{molec}} \xi_{r,s} n_s} \label{eq:lt-inner}, \\
    \left<E\right>_\text{tr}(T_{\text{tr,rel}}) &= &&\,\left<E\right>_\text{tr}(T_{\text{tr}}) - \sum_{r}\sum_{s}^{M_\text{molec}} \frac{\tau}{Z_{r,s}\tau_{s,\text{C}}}\left(	\left<E\right>_{r,s}(T_{\text{tr}}) - \left<E\right>_{r,s}(T_{r,s})  \right) \nonumber \\
    & &&\,- \tau \Delta H \frac{3 n}{3 n + \sum_r\sum_{s}^{M_\text{molec}} \xi_{r,s} n_s} \label{eq:lt-tr} 
\end{alignat}
with $r = \text{rot, vib}$.
$Z_{r,s}$ is the collision number of the internal degrees of freedom, $\tau_{s,\mathrm{C}}=1/\nu_{s,\mathrm{C}}$ is the collision time for the variable hard sphere model (VHS)~\cite{chapman-cowling}, and $\tau=1/\nu$ is the relaxation time.
Note that $\nu \neq \nu_{s,\mathrm{C}}$.

\section{Implementation}
\label{sec:implementation}
The proposed ESBGK chemistry model is implemented in the open-source code PICLas~\cite{piclas,piclas-bgk}.
The current implementation features diatomic molecules, while the extension to polyatomic molecules is planned for future work.
The particles in a cell relax with a probability
\begin{equation}
    P = 1 - \exp\left(-\nu \Delta t\right)
\end{equation}
in each time step $\Delta t$ towards their respective target distribution function $f_s^\text{ESBGK}$.
A subset of particles is found for the relaxation by comparing $P$ to an equally distributed random number $R_p\in[0,1)$ for each particle.
From this, the particles participating in a chemical reaction can be selected.

\subsection{Solving of rate equations and particle conversion}
For the particle method implementation, a change of the particle count for a specific reaction $r$ and species $s$ is derived from Eq.~\eqref{eq:dens-change} with the particle weight $w$, the time step $\Delta t$, and the volume $V$:
\begin{equation}
    \Delta N_{r,s}^* = \frac{\partial n_{r,s}}{\partial t} \frac{V \Delta t}{w}.\label{eq:number-change}
\end{equation}
If two or more reactions occur in parallel, a coupled differential equation system needs to be solved due to the changing particle number densities.
The particle change $\Delta N_{r,s}$ per species and reaction from Eq.~\eqref{eq:number-change} is subsequently obtained by a conversion to integer values using a random number $R_{r,s}\in [0,1)$
\begin{equation}
	\Delta N_{r,s} = \lfloor \Delta N_{r,s}^* + R_{r,s} \rfloor.
\end{equation}
This determines the number of particles that need to be added to or removed from the respective cell.
To ensure element and mass conservation within each reaction, this is only done for the particle with the smallest stoichiometric coefficient for each reaction, whereas the changes for all other species are calculated based on this to match the reaction equation.
The respective number of particles of the reactants are deleted randomly from the relaxation subset and the corresponding particles of the products are added subsequently.
The internal and relative kinetic energies of all reactants are summed up to $E_\text{old}$ for the energy conservation step, which is described in greater detail later.
The new relaxation subset now consists of all relaxing particles that are not involved in a reaction and the products generated in the reactions.

\subsection{New particle states}
For all particles in the new relaxation subset, the new particle states are determined by sampling from the target distribution function $f_s^\text{ESBGK}$.
The particle velocities are determined by a transformation of a random vector $\mathbf{r}_p$ chosen from a Maxwellian to a velocity vector chosen from the ESBGK distribution~\cite{gallis-torczynski}:
\begin{equation}
	\mathbf{v}_p^* = \mathbf{u} + \sqrt{\frac{k_\text{B} T_\text{tr,rel}}{m_s}} \left[ \mathbf{I} - \frac{1-\alpha Pr}{2\alpha Pr}  \left( \frac{m_s}{k_\text{B} T_\text{tr}} \mathbf{\Theta} - \mathbf{I} \right) \right] \mathbf{r}_p.
\end{equation}
The new internal energies of the relaxing particles are calculated by
\begin{align}
	E_{\text{rot},p}^* &= - \text{ln}(R_{p,1}) \frac{\xi_{\text{rot},s}}{2} k_\text{B} T_{\text{rot,rel},s},\\
    E_{\text{vib},p}^* &= - \text{ln}(R_{p,2}) \frac{\xi_{\text{vib},s}}{2} k_\text{B} T_{\text{vib,rel},s}
\end{align}
using equally distributed random numbers $R_{p,1}$ and $R_{p,2}$.
A rigid rotor model is used for the rotation, while a Simple Harmonic Oscillator model (SHO) is used for the vibration.
The zero-point energy is considered at the end of the time step.
These post-relaxation energies, hereby denoted with $^*$, still need to be scaled in a energy conservation step, which is described in the next section.

\subsection{Energy and momentum conservation}
Due to the stochastic nature of the particle method, the energy and momentum of the particles in a cell are not automatically conserved after the relaxation and reaction step, and a large number of particles would be needed for stable simulations.
Therefore, scaling factors are introduced to ensure energy and momentum conservation.
This is done based on the mean energies according to the Landau-Teller relaxation (see Eqs.~\eqref{eq:lt-inner}, \eqref{eq:lt-tr}).
Final particle velocities and internal energies are indicated with a prime $'$.

First, the vibrational energies are scaled due to the possibility of a quantized treatment.
For continuous vibrational energies, the scaling is
\begin{align}
    \alpha_{\text{vib},s} &= \frac{\left<E\right>_{\text{vib},s}(T_{\text{vib,rel},s}) N_{\text{rel},s}}{\sum_{p=1}^{N_{\text{rel},s}} E^*_{\text{vib},p,s}}, \\
    E_{\text{vib},p}' &= \alpha_{\text{vib},s} E_{\text{vib},p}^* + \frac{1}{2} k_\text{B} \Theta_{\text{vib},s}.
\end{align}
For quantized vibrational energies, a reformulation to a quantum number is necessary.
The final vibrational energies are subsequently subtracted from $E_\text{old}$.

With the remaining energy, the translational and rotational scaling factors can be calculated:
\begin{alignat}{2}
    \alpha_{\text{rot},s} &= \; &&\frac{E_\text{old}}{E_{\text{tr,not-rel}} + \left<E\right>_\text{tr}(T_{\text{tr,rel}}) N_\text{rel} + \sum_s \left<E\right>_{\text{rot},s}(T_{\text{rot,rel},s}) N_{\text{rel},s}} \nonumber \\
    & &&\cdot \frac{\left<E\right>_{\text{rot},s}(T_{\text{rot,rel},s}) N_{\text{rel},s}}{\sum_{p=1}^{N_{\text{rel},s}} E^*_{\text{rot},p,s}}, \\
    \alpha_{\text{tr}} &= \; &&\left[ \frac{E_\text{old}}{E_{\text{tr,not-rel}} + \left<E\right>_\text{tr}(T_{\text{tr,rel}}) N_\text{rel} + \sum_s \left<E\right>_{\text{rot},s}(T_{\text{rot,rel},s}) N_{\text{rel},s}} \right.\nonumber \\
    & &&\cdot \left. \frac{E_{\text{tr,not-rel}} + \left<E\right>_\text{tr}(T_{\text{tr,rel}}) N_\text{rel}}{\sum_{p=1}^{N_\text{rel}} \frac{m_s}{2} \left|\mathbf{v}_p^*-\mathbf{u}^*\right|^2} \right]^{1/2}.
\end{alignat}
Hereby, $E_{\text{tr,not-rel}}$ is the translational energy of the non-relaxing particles relative to the bulk velocity.
Thus, the final particle states are calculated as
\begin{align}
    \mathbf{v}_p' &= \mathbf{u} + \alpha_{\text{tr}} \left(\mathbf{v}_p^* - \mathbf{u}^*\right), \\
    E_{\text{rot},p}' &= \alpha_{\text{rot},s} E_{\text{rot},p}^*.
\end{align}

\section{Results}
\label{sec:results}

Two different test cases of a reactive gas confined in a reservoir are presented to demonstrate the functionality of the ESBGK model and compare its accuracy to DSMC results.
The first test case is a NO dissociation reaction
\begin{equation}
    2 \text{ NO} \rightarrow \text{N} + \text{O} + \text{NO}
\end{equation}
with the initial conditions $T_0 = \SI{20000}{\kelvin}$ and $n_{\text{NO},0} = \SI{1e22}{\per\cubic\meter}$.
The second test case is an exchange reaction with $T_0 = \SI{20000}{\kelvin}$, $n_{s,0} = \SI{1e22}{\per\cubic\meter}$:
\begin{equation}
    \text{N}_2 + \text{O} \rightarrow \text{NO} + \text{N}.
\end{equation}

\begin{figure}
  \centering
  \begin{subfigure}[b]{0.48\textwidth}
    \centering
    \includegraphics{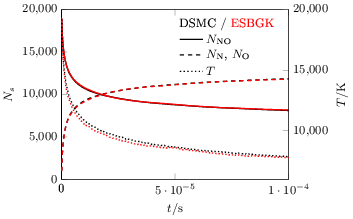}
    \vspace{-5mm}
    \caption{Dissociation: 2 NO $\rightarrow$ N + O + NO}
    \label{fig:diss}
  \end{subfigure}
  \hfill
  \begin{subfigure}[b]{0.48\textwidth}
    \centering
    \includegraphics{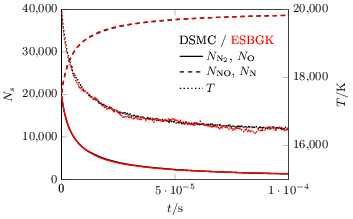}
    \vspace{-5mm}
    \caption{Exchange: N\textsubscript{2} + O $\rightarrow$ NO + N}
    \label{fig:exch}
  \end{subfigure}
  \vspace{-2mm}
  \caption{Simulation particle number and temperature over time.}
  \label{fig:tests}
\end{figure}

Figure~\ref{fig:diss} compares the number of simulation particles of all species participating in the dissociation reaction, as well as the temperature for this reaction over time.
The ESBGK method shows very good agreement with the DSMC result.
Figure~\ref{fig:exch} similarly compares the number of simulation particles and the temperature over time for the exchange reaction, where the ESBGK method shows good agreement with the DSMC result as well, demonstrating the accuracy of the model for modeling chemical reactions in simple gas mixtures.

\vspace{-5mm}

\section{Conclusion}
\label{sec:conclusion}
An Ellipsoidal Statistical BGK method for chemically reactive diatomic gas mixtures is proposed and implemented in the open-source particle code PICLas based on different BGK models and previous implementations~\cite{MATHIAUD202265,energy-cons,brull,brull2021ellipsoidal,bgk-polymulti,bgk-chem}.
As test cases for verification, different gases confined in a reservoir, including a dissociation and an exchange reaction, are presented, where overall very good agreement with the DSMC method is shown.
In future work, the extension to polyatomic molecules is envisioned, and the model will be applied to the Shakhov BGK model~\cite{shakhov}.
Furthermore, the handling of dissociation reactions with the Quantum Kinetics approach will be investigated as an alternative.

\begin{credits}
\subsubsection{\ackname} This work is funded by the Deutsche Forschungsgemeinschaft (DFG, German Research Foundation) -- Project-ID 516238647 -- SFB 1667 (ATLAS - Advancing Technologies of Low-Altitude Satellites).

\subsubsection{\discintname}
The authors have no competing interests to declare that are relevant to the content of this article.
\end{credits}

%
%
%

\begin{thebibliography}{30}

\bibitem {bird}
Bird, G.A.: Molecular Gas Dynamics and the Direct Simulation of Gas Flows. Oxford University Press, New York (1994)

\bibitem {bgk}
Bhatnagar, P.L., Gross, E.P., Krook, M.: A Model for Collision Processes in Gases. I. Small Amplitude Processes in Charged and Neutral One-Component Systems. Phys. Rev. 94(3), 511--525 (1954)

\bibitem {esbgk}
Holway Jr., H.L.: New Statistical Models for Kinetic Theory: Methods of Construction. Phys. Fluids 9, 1658--1673 (1966)

\bibitem {shakhov}
Shakhov, E.M.: Generalization of the Krook Kinetic Relaxation Equation. Fluid Dyn. 33, 95--96 (1968)

\bibitem {fp-basic}
Jenny, P., Torrilhon, M., Heinz, S.: A solution algorithm for the fluid dynamic equations based on a stochastic model for molecular motion. J. Comput. Phys. 229(4), 1077--1098 (2010)

\bibitem {piclas}
Fasoulas, S., Munz, C.-D., Pfeiffer, M., Beyer, J., Binder, T., Copplestone, S., Mirza, A., Nizenkov, P., Ortwein, P., Reschke, W.: Combining particle-in-cell and direct simulation Monte Carlo for the simulation of reactive plasma flows. Phys. Fluids 31, 072006 (2019)

\bibitem {piclas-bgk}
Pfeiffer, M.: Particle-based fluid dynamics: Comparison of different Bhatnagar-Gross-Krook models and the direct simulation Monte Carlo method for hypersonic flows. Phys. Fluids 30, 106106 (2018)

\bibitem {bgk-polymulti}
Hild, F., Pfeiffer, M.: Multi-species modeling in the particle-based ellipsoidal statistical Bhatnagar-Gross-Krook method including internal degrees of freedom. J. Comput. Phys. 514, 113226 (2024)

\bibitem {bgk-chem}
Tuttas, F., Pfeiffer, M.: Modeling of particle-based BGK methods for the simulation of chemically reactive gas mixtures in multi-scale flows. In: 3rd International Conference on Flight vehicles, Aerothermodynamics and Re-entry, Arcachon, France (2025)

\bibitem {ATcT}
Ruscic, B., Pinzon, R. E., von Laszewski, G., Kodeboyina, D., Burcat, A., Leahy, D., Montoya, D., Wagner, A. F.: Active Thermochemical Tables: thermochemistry for the 21st century. J. Phys.: Conf. Ser. 16, pp. 561 (2005)

\bibitem {MATHIAUD202265}
Mathiaud, J., Mieussens, L., Pfeiffer, M.: An ES-BGK model for diatomic gases with correct relaxation rates for internal energies. Eur. J. Mech. B/Fluids 96, 65--77 (2022)

\bibitem {dauvois2021bgk}
Dauvois, Y., Mathiaud, J., Mieussens, L.: An ES-BGK model for polyatomic gases in rotational and vibrational nonequilibrium. Eur. J. of Mech. B/Fluids 88, 1--16 (2021)

\bibitem {energy-cons}
Pfeiffer, M.: Extending the particle ellipsoidal statistical Bhatnagar-Gross-Krook method to diatomic molecules including quantized vibrational energies. Phys. Fluids 30, 116103 (2018)

\bibitem {brull}
Brull, S.: An ellipsoidal statistical model for gas mixtures. Commun. Math. Sci. 13, 1--13 (2014)

\bibitem {brull2021ellipsoidal}
Brull, S.: An ellipsoidal statistical model for a monoatomic and a polyatomic gas mixture. Commun. Math. Sci. 19(8), 2177--2194 (2021)

\bibitem {chapman-cowling}
Chapman, S., Cowling, T. G.: The Mathematical Theory of Non-uniform Gases, Cambridge University Press (1970)

\bibitem {collint}
Hirschfelder, J.O., Curtiss, C.F., Bird, R.B.: The Molecular Theory of Gases and Liquids. Wiley-Interscience (1964)

\bibitem {gallis-torczynski}
Gallis, M.A., Torczynski, J.R.: The Application of the BGK Model in Particle Simulations. In: 34th Thermophysics Conference, American Institute of Aeronautics and Astronautics, Denver, Colorado, USA (2000)

\end{thebibliography}
%

\end{document}